\documentclass[lettersize,journal]{IEEEtran}
\usepackage{stfloats}
\usepackage{amsfonts}
\usepackage{amssymb}
\usepackage{amsthm}
\usepackage{cite}
\usepackage[cmex10]{amsmath}
\usepackage{float}
\usepackage{color}
\usepackage{multirow}
\usepackage[normalem]{ulem}
\usepackage{tabularx}
\usepackage{subfigure}
\usepackage{fancybox,dashbox}
\usepackage{bbm}
\usepackage[dvipsnames]{xcolor}
\usepackage{balance}
\usepackage{algorithm}
\usepackage{algpseudocode}
\usepackage{enumitem}
\usepackage{bm}
\usepackage{geometry}
\ifCLASSINFOpdf
\usepackage[pdftex]{graphicx}
\DeclareGraphicsExtensions{.pdf,.jpeg,.png}
\else
\usepackage[dvips]{graphicx}
\DeclareGraphicsExtensions{.eps}
\fi

\begin{document}

\title{
Dynamic XR Rendering Offloading Based on Feature-Based Quality Assessment}

\author{Sige Liu,~\IEEEmembership{Member,~IEEE,}  Zhe Wang, Lavish Kamal Kumar, and Yansha Deng,~\IEEEmembership{Senior Member,~IEEE}
\thanks{S. Liu, Z. Wang, L. K. Kumar, and Y. Deng are with the Department of Engineering, King's College London, London WC2R 2L, U.K. (e-mail: sige.liu@kcl.ac.uk; tylor.wang@kcl.ac.uk; 
lavishkk08@gmail.com; yansha.deng@kcl.ac.uk) (Corresponding auther: Yansha Deng).} 
\thanks{Demonstration is available online at https://www.youtube.com/watch?v=iv
w\_i8LXmi8.} \vspace{-15pt}
}



\maketitle

\begin{abstract}

Extended Reality (XR) applications demand intensive computation and low latency, especially for real-time rendering tasks. In this letter, we present an edge-aided XR rendering testbed that dynamically offloads rendering workloads between the XR client and the edge server built upon network conditions and latency constraints. The testbed integrates a Microsoft HoloLens 2 headset, a GPU-enabled edge server, and a customized remote rendering toolkit based on the HOLO Stream SDK, enabling seamless switching between local and edge rendering modes in real time.
To overcome the limitations of pixel-level quality metrics under head movements and asynchronous frame arrivals, we propose a perceptual evaluation metric based on deep feature embeddings and cosine similarity, which remains robust to spatial and temporal misalignments.
Furthermore, we design a contextual bandit learning controller to adapt rendering placement decisions in real time by jointly optimizing perceptual quality and latency.
Experimental results demonstrate the feasibility and performance of our testbed, validating its effectiveness in delivering high-quality and interactive XR experiences.
\end{abstract}


\section{Introduction}

Extended Reality (XR), encompassing Augmented Reality (AR), Virtual Reality (VR), and Mixed Reality (MR), is rapidly transforming various verticals such as remote maintenance, immersive healthcare, smart manufacturing, and collaborative digital twins \cite{cao2023exploring}. These applications demand real-time interaction, low-latency communication, and high-fidelity 3D rendering, especially in mobile environments \cite{malandrino2025xr}. However, the limited onboard computational capacity and energy constraints of these lightweight devices make it challenging to meet the rendering requirements of modern XR computation workloads \cite{nagy2023ai}.

To overcome these limitations, a promising solution lies in offloading computationally intensive rendering tasks to nearby edge servers. In such architectures, the camera at the XR side sends the captured 2D video to the edge server, which processes and returns rendered 3D content. While this paradigm can significantly enhance rendering quality and reduce on-device computation, it also introduces a real-time control dilemma: where and when should rendering be executed to balance latency, quality, and system reliability \cite{pan2024quality}?
Offloading decision becomes non-trivial under dynamic wireless conditions. Local rendering ensures low-latency response but is constrained by limited graphics processing power of XR devices, often resulting in poor visual fidelity and incomplete scene reconstruction \cite{nyamtiga2024adaptive}. In contrast, edge rendering delivers high-quality output but is sensitive to fluctuating network bandwidth and latency, which can result in perceptible lag, degraded immersion, or even failures \cite{chukhno2025application}. Therefore, a robust and adaptive offloading mechanism is required to dynamically balance latency constraints and perceptual quality.


A further challenge is to assess and control XR rendering quality when head motion and network jitter break pixel-level alignment and temporal coherence \cite{mantiuk2024colorvideovdp}. Pixel-wise metrics (e.g., PSNR/SSIM) and sequence-coherent scores (e.g., FVD) become unreliable under such misalignments, which undermines any offloading policy that depends on quality feedback \cite{wu2023assessor360}. Meanwhile, most existing research assumes stable links or remains simulation-only \cite{heo2023flexr}\cite{morin2023extended}, lacking a deployable framework that 1) measures perceptual quality robustly online, 2) adapts rendering offloading to the latency–quality trade-off, and 3) runs edge-aided on real devices and networks. 
To address this gap, we expose the real-time video quality metric as feedback from the XR user to the edge server, and propose a learning-based dynamic rendering offloading based on the feedback metric, taking into account the latency constraints, and a practical testbed to execute and validate decisions in real networks.

We build an edge-aided XR offloading testbed with a HoloLens-2 client and a GPU edge server connected via HOLO Stream\footnote{HOLO Stream: https://github.com/Holo-Light-GmbH/Hololight-Stream-SDK-Trial.} and containerised orchestration, enabling seamless migration between local and edge rendering. 
To better quantify visual quality robustly in asynchronous rendering, we leverage the Visual Geometry Group (VGG) network as a pre-trained convolutional feature extractor and compute cosine similarity between its clip-level embeddings \cite{jindal2025cgvqm}. That is because these high-level features exhibit greater robustness to spatial and temporal misalignments compared with pixel-based metrics. 
For the stream part, we integrate HOLO Stream into a custom control and data plane that exposes a binary flag for mode switching and continuously measures end-to-end latency and throughput. 
The edge monitors these indicators and updates the flag, while the client periodically polls it to switch rendering modes dynamically. 
Building on this feedback loop, a contextual bandit learning controller utilizes the VGG quality signal together with the latency to select the rendering location (client vs. edge), and the testbed executes the decision in real time.
Compared to all-edge or all-client rendering, our approach demonstrates better adaptability to changing network conditions and latency requirements, improving user-perceived quality and reducing end-to-end latency in a practical edge-aided XR testbed.
Our contributions are summarized as: 
\begin{itemize}
    \item We develop an edge-aided testbed for dynamic offloading XR/edge switching and real-network validation, integrating HoloLens-2, HOLO Stream, and a GPU edge with containerised orchestration.
    \item We introduce an online perceptual quality measure based on VGG feature embeddings with cosine similarity, providing robustness to spatial and temporal frame misalignments in XR video evaluation.
    \item We propose a contextual-bandit offloading policy that optimises visual quality under latency constraints, offering a lightweight and deployable solution for XR devices
\end{itemize}
The rest of this letter is organised as follows. Section~II presents the system overview. Section~III describes the architecture and implementation. Section~IV introduces the offloading formulation and learning-based controller. Section~V reports the experimental results. Finally, Section~VI concludes the letter.

\section{System Overview}
We develop an edge-aided testbed to support dynamic XR rendering task offloading between XR and edge. The testbed integrates real-time video collection, wireless transmission, rendering orchestration, and perceptual quality evaluation. This section outlines both the hardware setup and software architecture.

\subsection{Hardware}

As shown in Fig.~\ref{fig:hardware}, our XR testbed provides an edge-aided framework that supports both client-side and edge-side rendering, and it consists of three primary components:

\begin{itemize}
  \item Client Device: A Microsoft HoloLens 2 headset functions as the XR client, capable of switching between local rendering and edge streaming via the HOLO stream.
  \item Edge Server: A GPU-enabled workstation equipped with Unity and HOLO Stream, deployed in Docker containers to support scalable rendering tasks.
  \item Video Source: 2D video feed (e.g., from a drone camera or pre-recorded source) providing the scene content that is rendered either locally at the XR or remotely at the edge server.
\end{itemize}




\subsection{Software}

\begin{figure}[t]
  \centering
  \includegraphics[width=0.9\linewidth]{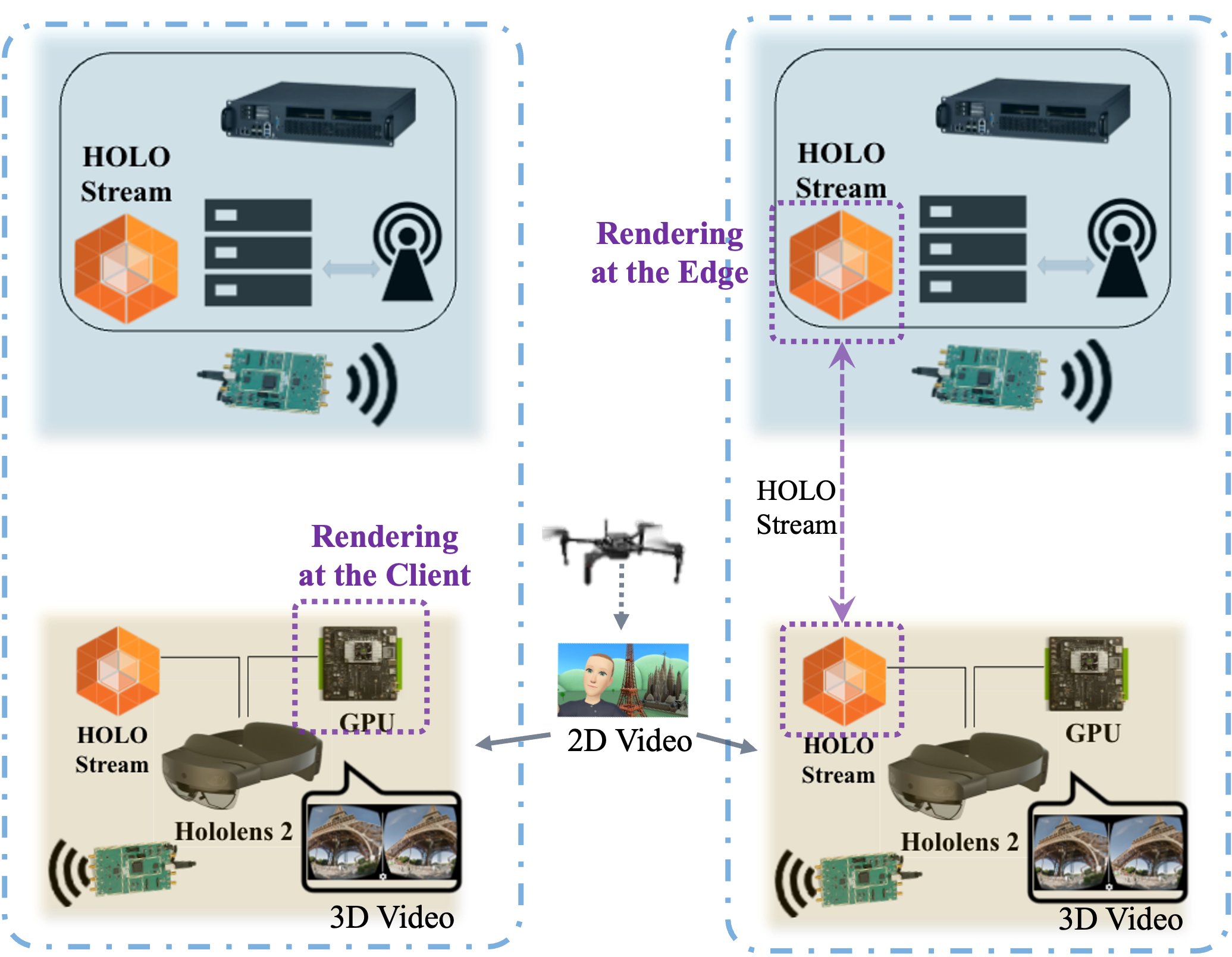}
  \caption{The framework of the dynamic XR rendering testbed.}
  \label{fig:hardware}
\end{figure}

At a high level, the software stack follows Fig.~\ref{fig:hardware} and decouples streaming, rendering, and control. 
HOLO Stream provides the remote-rendering SDK used for bidirectional streaming between client and edge, including frame encoding, transport, decoding, and forwarding. We integrate it with a control plane for seamless mode switching. A high-level overview of the software is as follows.

\begin{itemize}
    \item Unity Rendering: the XR scene runs in Unity on both ends: the HoloLens~2 executes the scene at the local side, whereas the GPU server renders the scene at the edge side.

    \item HOLO Stream: the HOLO Stream client/server pair handles remote-rendering streaming, i.e., delivering frames from the active renderer (client or edge) to the XR and forwarding user inputs back to the renderer, which enables the two rendering modes.

    \item Dynamic Offloading: it coordinates client and edge rendering and ensures seamless switching during the XR session without interrupting the pipeline. The detailed design and our modifications are presented in Section~III.

\end{itemize}

\section{Dynamic XR Offloading Architecture}
This section details our modifications to the HOLO Stream and the runtime control that enable seamless migration between client-side and edge-side rendering. We first describe how we integrate and extend HOLO Stream, and then present the dynamic rendering process at the server and client.

\subsection{Integrating and Modifying HOLO Stream}

HOLO Stream provides the SDK for remote rendering, and we integrate it into our pipeline and extend it with a control plane that exposes mode control and runtime telemetry to the decision module. The reference implementation and scripts are available on our repository.\footnote{The Docker API: https://github.com/GetupOscar/Poll-API}

As shown in Fig.~2, we introduce an interface at the edge that maintains a binary RenderFlag indicating the target rendering location (\texttt{0}=edge, \texttt{1}=client). The XR client dynamically polls this flag over HTTPS at a fixed interval and switches modes without interrupting the session. This indirection decouples decision logic from the rendering pipeline and allows us to plug in either threshold rules or learning-based policies with no changes to streaming.

We add a profiling hook that continuously collects end-to-end latency and throughput. These measurements are made available to the controller and also logged for offline analysis. Safeguard thresholds $(\tau_L,\tau_B)$ can be applied to enforce fallbacks when latency is excessive or bitrate is insufficient.
At the edge, Unity renderers are launched and supervised as services; on mode changes, the streaming channel is re-bound to the active renderer while keeping the XR session alive at the client.

\subsection{XR-Side Control}
The XR runs a lightweight controller that periodically polls the RenderFlag from the edge server. If the flag is \texttt{0}, the client connects to the edge HOLO Stream server and displays remotely rendered frames; if the flag is \texttt{1}, it switches to local Unity rendering. The polling is non-blocking and includes time-out handling to avoid user-perceived glitches. This client-side state machine performs transparent switching during an active session and corresponds to \textbf{Algorithm~1}.

\subsection{Edge-Side Orchestration}

The edge server hosts the Unity renderer and the HOLO~Stream service and orchestrates remote rendering on a GPU workstation. Its responsibilities are to expose runtime latency and mange switch mode to ensure seamless migration between client and edge.

At runtime, the edge continuously monitors end-to-end latency $L_t$ and throughput $B_t$ from the active session and makes them available to the controller. The controller’s output is realised through a binary mode indicator RenderFlag $\in\{0,1\}$. In addition, safeguard thresholds $(\tau_L,\tau_B)$ are enforced to guarantee responsiveness: if $L_t>\tau_L$ or $B_t<\tau_B$, the edge immediately issues a fallback to client mode.
When switching into edge mode, the orchestration layer ensures the Unity renderer is available before rebinding the HOLO~Stream channel, thereby avoiding visible glitches. When leaving edge mode, remote rendering is suspended and session state remains consistent for an instantaneous return to client rendering. The renderer runs as a managed service, enabling clean start/stop operations and recovery on failure. The general logic is summarized in \textbf{Algorithm~2}.

\begin{figure}[t]
  \centering
  \includegraphics[width=0.45\linewidth]{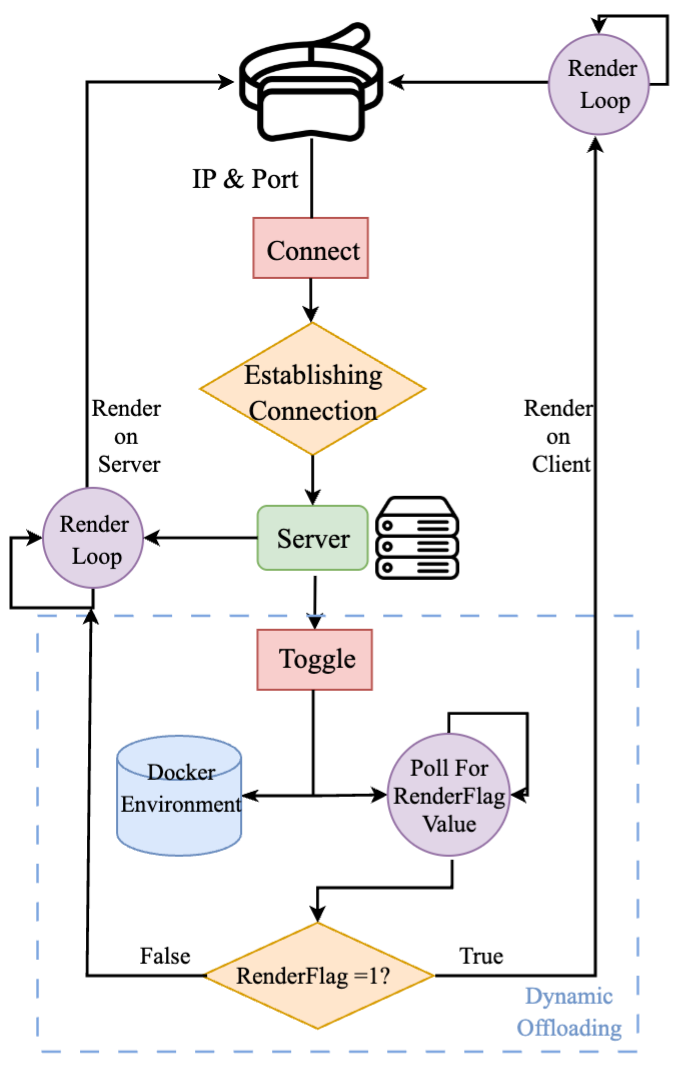}
  \caption{ Workflow of the dynamic offloading and control mechanism.}
  \label{fig:software}\vspace{-5pt}
\end{figure}
\begin{algorithm}[h]
\caption{Client-Side Rendering Control}
\begin{algorithmic}[1]
\State Initialize XR session on HoloLens.
\State Set polling interval $T_{\text{poll}}$.
\While{XR session is active}
    \State Send HTTPS GET request to RenderFlag.
    \State Receive flag $F \in \{0, 1\}$ ($1{=}$Client, $0{=}$Serve).
    \If{$F = 0$}
        \State Switch to Server Rendering Mode.
        \State Connect to edge server via HOLO Stream.
        \State Stream videos for display.
    \Else
        \State Switch to Client Rendering Mode.
        \State Launch Unity scene on-device.
    \EndIf
\EndWhile
\end{algorithmic}
\end{algorithm}
\begin{algorithm}[h]
\caption{Edge-Side Orchestration}
\begin{algorithmic}[1]
\State Initialise Docker environment..
\State Set safeguard thresholds $\tau_L,\tau_B$.
\While{server is active}
    \If{latency $> \tau_L$ or bitrate $< \tau_B$}
        \State Set RenderFlag $\leftarrow 1$ \Comment{Client rendering}
        \If{container is running}
            \State Stop the container and release resources..
        \EndIf
    \Else
        \State Set RenderFlag $\leftarrow 0$ \Comment{Edge rendering}
        \If{container not running}
            \State Launch Unity rendering container.
        \EndIf
        \State Receive camera stream and render in Unity.
        \State Stream output to HoloLens 2.
    \EndIf
\EndWhile
\end{algorithmic}
\end{algorithm}


\subsection{Problem Formulation}
The core objective of our dynamic XR offloading system is to maximize perceptual rendering quality while meeting strict end-to-end latency constraints imposed by the user experience. Given a set of rendering modes $\mathcal{A} = \{0: \text{local}, 1: \text{edge}\}$ and a sequence of rendering decisions over time steps $t = 1, \dots, T$, we aim to select an action $A_t \in \mathcal{A}$ that balances the trade-off between visual quality and latency.

We denote the perceptual similarity between the rendered clip and the reference clip at time $t$ by $\text{V}_t$, computed via VGG feature cosine similarity. Let $L_t$ denote the observed latency, which includes rendering time, network transmission delay, and any feature extraction overhead.
The system's objective is to find a policy $\pi$ that maps context information to actions, such that the expected cumulative utility over time is maximized:
\begin{equation}
\mathcal{P:} \quad \max_{\pi} \; \mathbb{E} \left[ \sum_{t=1}^T V_t \right], \quad \text{s.t.} \; L_t \leq L^{\text{req}}_t \; \forall t
\end{equation}
where $L^{\text{req}}_t$ denotes the latency requirement specified by the user at time step $t$, and this formulation captures the trade-off between perceptual quality and responsiveness, ensuring that the selected rendering policy remains aligned with user experience constraints.

\section{Learning-Based Offloading Decision}

This optimization is challenging due to the dynamic wireless conditions and the absence of a priori statistical models.
To address these constraints, we adopt a contextual bandit learning method, where the decision at each time step is based on observable context features such as latency budget, throughput, and historical reward. This approach provides a lightweight, explainable, and data-efficient mechanism for adaptive offloading in heterogeneous network environments.
To enhance the system’s flexibility and intelligence, we implement a contextual bandit learning strategy that selects the rendering location based on both quality and latency metrics.

\subsection{Context Representation and Quality Metric}
In contextual bandit learning, each decision is made with respect to an observable context \cite{vsvihrova2025designing}. In our formulation, the context is defined as the user-specified latency requirement $L^{\text{req}}_t$ at time step $t$, which reflects the upper bound on tolerable end-to-end delay. This latency constraint implicitly encodes the user’s preference between quality and responsiveness, and provides a compact yet informative representation of environment dynamics.

To measure perceptual quality, we employ a feature-based similarity metric robust to alignment and motion artifacts. Let $\mathbf{f}_{t,i}$ denote the VGG-extracted feature vector of frame $i$ in clip $t$, and let $T$ denote the number of frames per clip. The average clip-level feature embedding is computed as
\begin{equation}
\overline{\mathbf{f}}_t = \frac{1}{T} \sum_{i=1}^{T} \mathbf{f}_{t,i}
\end{equation}

The cosine similarity between the rendered clip and the reference clip is given by
\begin{equation}
\text{V}_t = \frac{\overline{\mathbf{f}}_t^{(\text{rendered})} \cdot \overline{\mathbf{f}}_t^{(\text{ref})}}{\|\overline{\mathbf{f}}_t^{(\text{rendered})}\| \cdot \|\overline{\mathbf{f}}_t^{(\text{ref})}\|},
\end{equation}
which is defined as the reward function.

\subsection{Contextual Bandit Strategy}

We model the offloading decision process as a two-armed contextual Bandit problem with action space $\mathcal{A} = \{0: \text{Edge}, 1: \text{Client}\}$. At each round $t$, the agent receives a context $x_t = L^{\text{req}}_t$, selects an action $A_t$, observes the reward $R_t$, and updates its value estimates accordingly.
We implement an $\varepsilon$-greedy learning strategy with online reward averaging in \textbf{Algorithm 3}.

\begin{algorithm}[H]
\caption{Contextual Bandit Learning Offloading Policy}
\begin{algorithmic}[1]
\State Initialize $Q_0(x), Q_1(x) \leftarrow 0$, $N_0, N_1 \leftarrow 0$ for all contexts $x_t$.
\State Set exploration rate $\varepsilon$.
\For{each decision round $t$}
    \State Observe context $x_t = L^{\text{req}}_t$.
    \State With prob. $1 - \varepsilon$, select $A_t \leftarrow \arg\max_{a \in \mathcal{A}} Q_a(x_t)$.
    \State With prob. $\varepsilon$, choose $A_t$ randomly.
    \State Execute rendering using mode $A_t$.
    \State Compute cosine similarity $V_t$ via Eq.~(3).
    \State Update:
    \begin{align*}
        N_{A_t}(x_t) &\leftarrow N_{A_t}(x_t) + 1. \\
        Q_{A_t}(x_t) &\leftarrow Q_{A_t}(x_t) + \frac{1}{N_{A_t}(x_t)} (R_t - Q_{A_t}(x_t)).
    \end{align*}
\EndFor
\end{algorithmic}
\end{algorithm}
This approach allows the agent to gradually learn the best action for each latency requirement while balancing exploration and exploitation. Unlike full RL methods, contextual Bandits require minimal computation and are better suited for resource-constrained XR platforms. Future extensions may consider richer contextual features or model-based Bandit variants.

\section{Simulation Results}
In this section, we evaluate our proposed XR rendering offloading framework via simulation. The primary objective is to investigate how the dynamic bandit-based offloading mechanism adapts to different latency requirements and balances the trade-off between rendering quality and latency.

\subsection{Experimental Settings}
To validate the effectiveness of our offloading testbed, we conduct a series of experiments using the deployed system under dynamic 5G network conditions. The XR client is a Microsoft HoloLens 2 headset running Unity, and the edge server is equipped with two NVIDIA A4000 GPUs. The experiments aim to evaluate the trade-off between rendering quality and latency constraints under different offloading strategies. 
We simulate 1,000 consecutive decision rounds. The system must decide, at each step, whether to offload the rendering task to the client or to the edge server. The video quality is quantified by cosine similarity between VGG-extracted features of the rendered clip and the reference ground truth. Higher similarity indicates better perceptual quality. The agent employs an $\epsilon$-greedy multi-armed bandit algorithm with $\epsilon = 0.05$, selecting the rendering mode based on estimated expected rewards. The simulation is repeated under four latency constraints: 20 ms, 40 ms, 60 ms, and 80 ms.
We compare the performance of the following policies: 1) Dynamic Offloading: our proposed method; 2) All local: All tasks are rendered locally; 3) All edge: All tasks are rendered at the edge.

\subsection{Performance Evaluation}
\begin{figure*}[t!]
\setlength{\abovecaptionskip}{-0.01cm}
		 \centering
  \subfigure[Reward Comparison]{
		\includegraphics[width=0.31\textwidth]{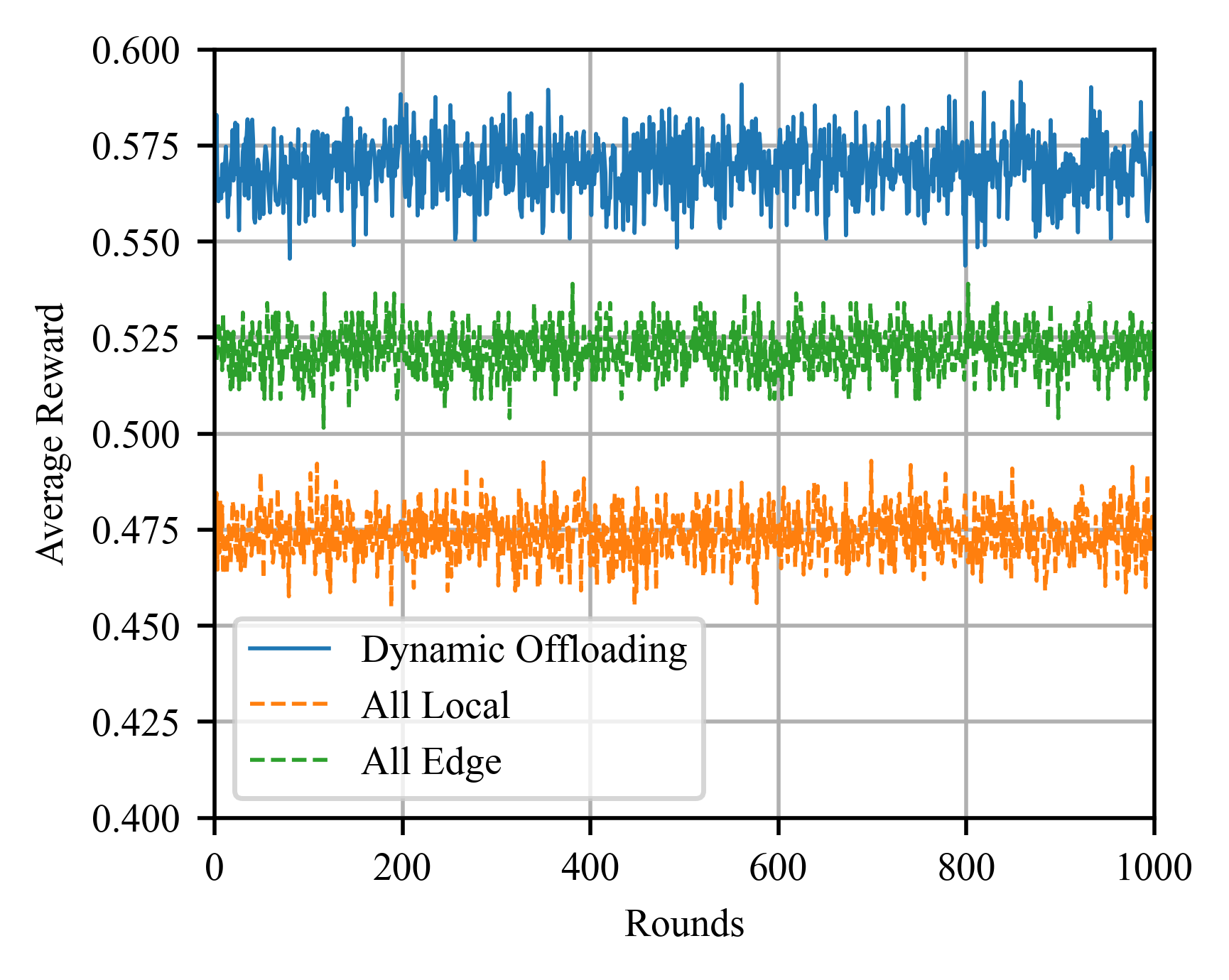}
		\label{simu_a}
		}
  \subfigure[Dynamic offloading under different constraints.]{
		\includegraphics[width=0.31\textwidth]{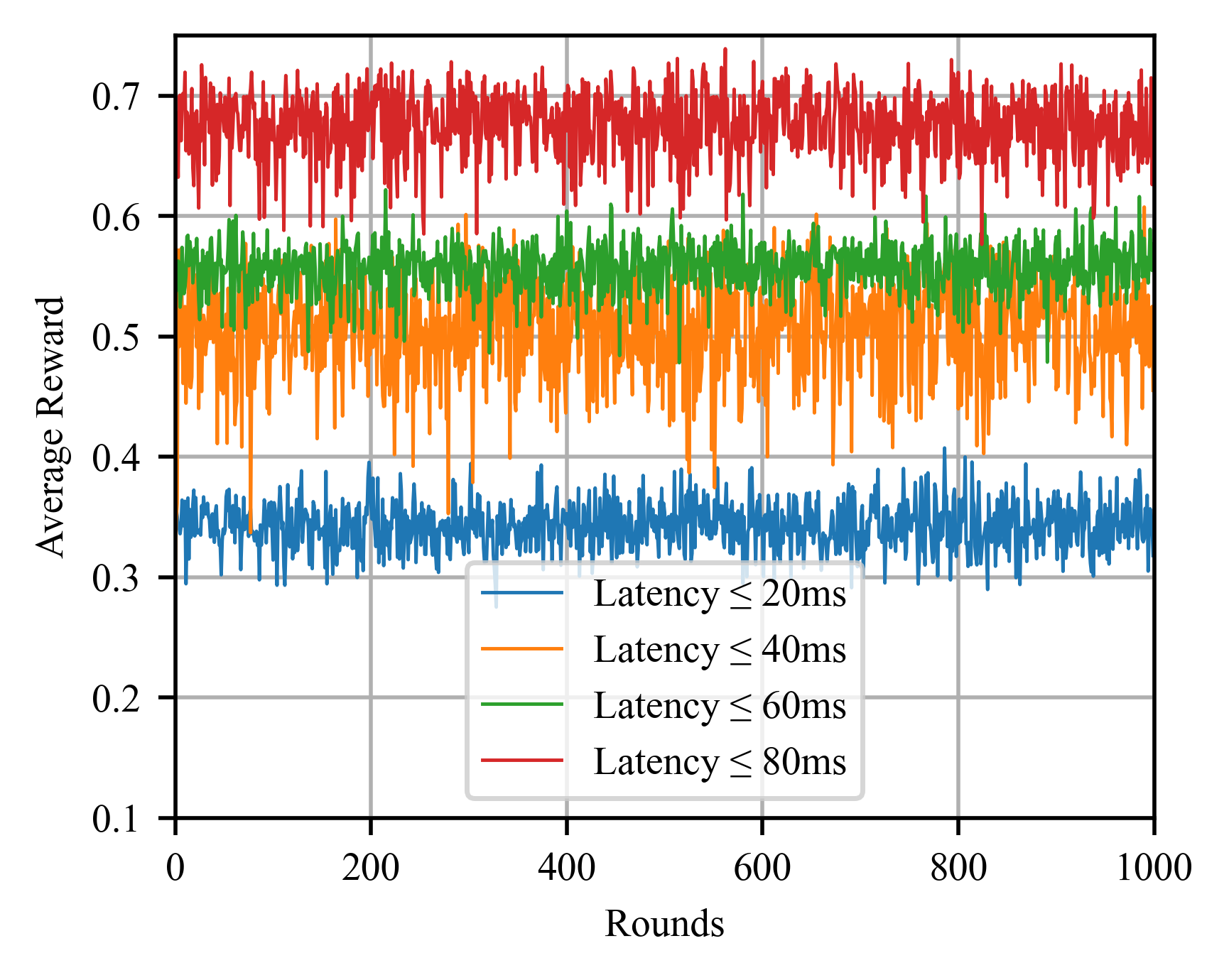}
		\label{simu_b}
		}
    \subfigure[ Action counts under different latency constraints]{
		\includegraphics[width=0.31\textwidth]{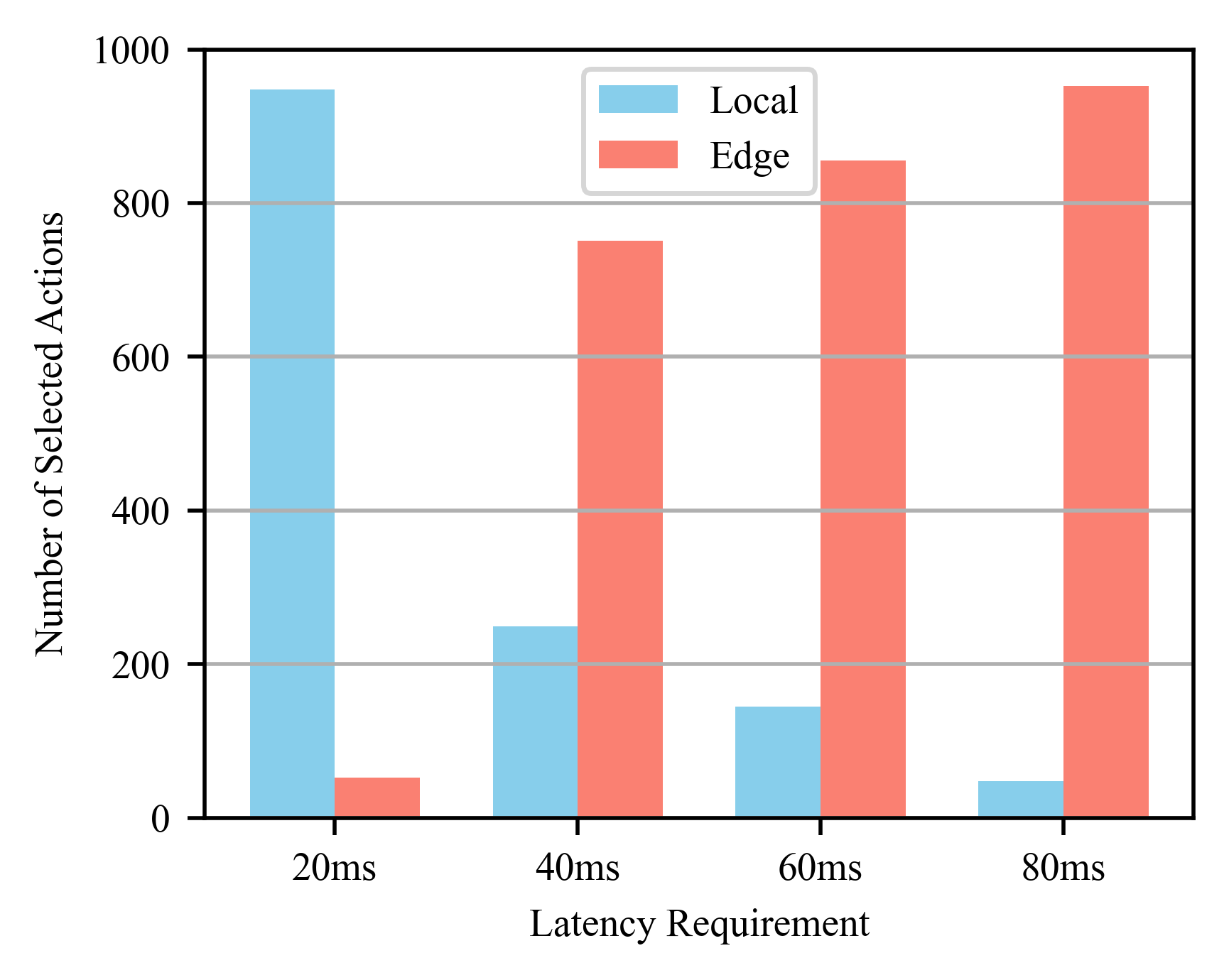}
		\label{simu_c}
		}
		\caption{The performance of the proposed dynamic offloading strategy.}
		\label{simu}
        \vspace{-5pt}
	\end{figure*}

Fig. \ref{simu_a} compares the average rewards achieved by the three strategies under a latency constraint of 40 ms. The dynamic offloading strategy consistently outperforms both baselines. The “All Local” method satisfies latency constraints but suffers from lower rendering quality. The “All Edge” approach provides higher quality but frequently violates the latency threshold, leading to penalized rewards. The proposed strategy balances these factors and achieves the highest stable reward across all rounds.
Fig. \ref{simu_b} illustrates the dynamic strategy's performance under different latency requirements. As the constraint becomes looser (from 20 ms to 80 ms), the average reward increases steadily. This trend demonstrates that our proposed method effectively adapts to the available latency budget, shifting its preference toward higher-quality edge rendering when the latency tolerance allows. The curves are stable, highlighting the robustness of the bandit agent in changing conditions.
Fig. \ref{simu_c} presents the total number of local and edge rendering actions selected under different latency requirements. Under tight constraints (e.g., 20 ms), local rendering is heavily favored to avoid latency violations. As the latency budget increases, the agent gradually shifts toward edge rendering to improve video quality. This clearly reflects the algorithm’s adaptive behavior and aligns with human-understandable offloading logic, contributing to the explainability of AI-driven decisions.

\begin{figure}[t]
		\centering
		\subfigure[Selected frame]{
	    \includegraphics[width=0.22\textwidth]{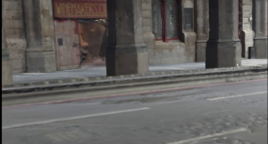}
		\label{selected frame}
		}
  \subfigure[VGG extracted features (first 8 channels)]{
		\includegraphics[width=0.22\textwidth]{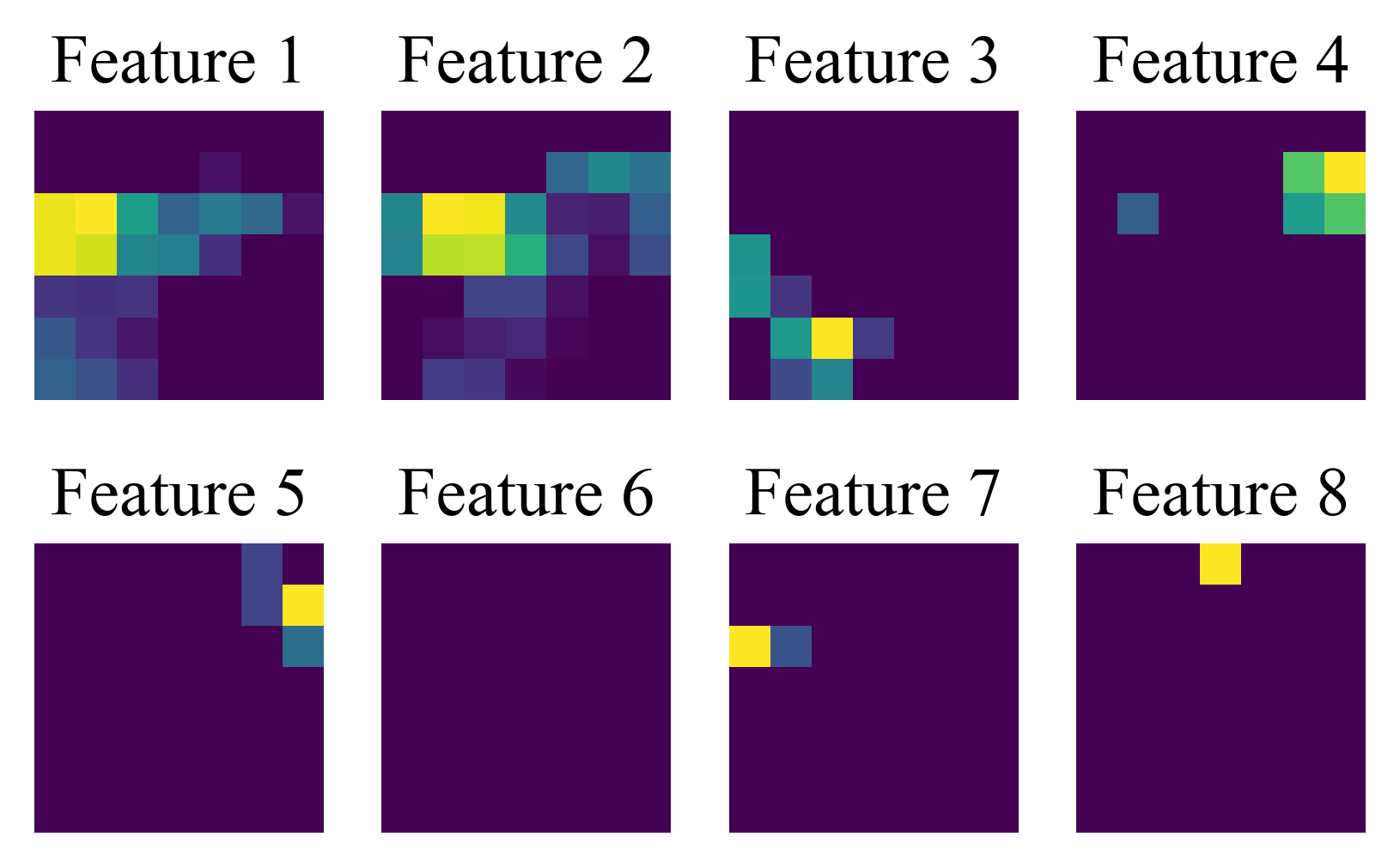}
		\label{vgg feature}
		}
		\caption{An example of a frame and its VGG extracted features.}
		\label{frame and vgg}
	\end{figure}

\begin{figure}[t]
		\centering
		\subfigure[Rendering at local device]{
	    \includegraphics[width=0.214\textwidth]{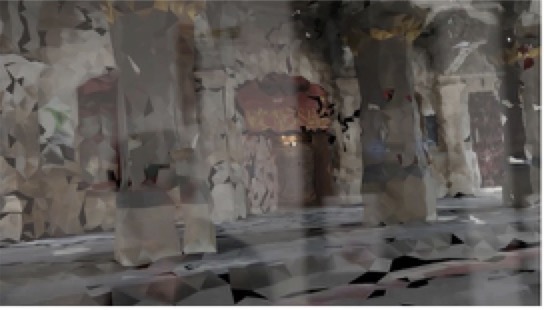}
		\label{render_a}
		}
  \subfigure[Rendering at the edge server]{
		\includegraphics[width=0.2\textwidth]{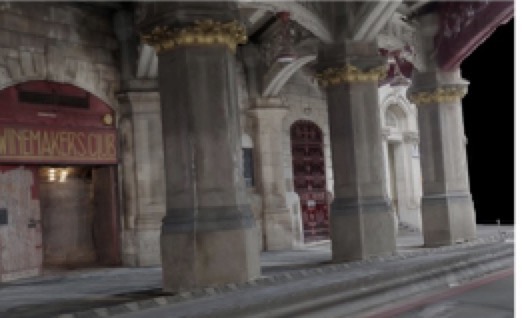}
		\label{render_b}
		}
		\caption{An example frame of comparison of rendering.}
		\label{render_all}
	\end{figure}

To provide intuitive insight into the perceptual difference between rendering options, we present two qualitative examples in Fig. \ref{frame and vgg} and Fig. \ref{render_all}.
Fig. \ref{frame and vgg} presents a selected frame from the ground truth video (Fig. \ref{selected frame}) alongside the corresponding VGG feature maps (Fig. \ref{vgg feature}) extracted from its first 8 channels. These features are used as semantic representations in the reward calculation pipeline, capturing global appearance patterns while remaining robust to pixel-level noise or temporal misalignment.
Fig. \ref{render_all} compares visual outputs rendered at the local device and edge server for the same frame. The locally rendered result (Fig. \ref{render_a} shows visible distortions, low-fidelity textures, and incomplete geometry, which degrade the immersive experience. In contrast, the edge-rendered frame (Fig. \ref{render_b}) preserves significantly more detail, showing sharp textures and a complete architectural structure. This highlights the benefit of edge rendering in scenarios where latency constraints permit higher computation latency in exchange for enhanced quality.


\vspace{-5pt}
\section{Conclusion}
In this letter, we developed an edge-aided XR offloading testbed that enables dynamic switching between XR and edge server rendering via a modified HOLO Stream stack and a lightweight control plane. A Visual Geometry Group (VGG)-embedding cosine-similarity metric provides alignment-robust feedback, and a contextual bandit learning uses this feedback together with latency to select the rendering location online. Experimental results and qualitative comparisons validate the effectiveness and adaptability of our framework under realistic conditions. Future work includes multi-user coordination and predictive rendering control.
\vspace{-10pt}
\bibliographystyle{IEEEtran}
\bibliography{Bibliography}

\end{document}